\documentstyle[aps,prl,multicol,epsf]{revtex}

\newcommand{\beq}{\begin{equation}}
\newcommand{\enq}{\end{equation}}
 
\begin{document}

\title{Hysteresis in driven disordered systems:
from plastic depinning to magnets} 
 
\author{M. Cristina Marchetti$^1$ and Karin A. Dahmen$^2$}
\address{$^1$Physics Department, Syracuse University, Syracuse, NY 13244}
\address{$^2$Loomis Laboratory, University of Illinois at Urbana-Champaign,
Urbana, IL 61801}
 
\maketitle

\begin{abstract}

We study the dynamics of a viscoelastic medium driven through quenched
disorder by expanding about mean field theory in $6-\epsilon$ dimensions.
The model exhibits a critical point separating a region where the dynamics
is hysteretic,
with a macroscopic jump between strongly pinned and
weakly pinned states,
from a region where the sliding state is unique
and no jump occurs. The disappearance of the jump at the critical
point is described by universal exponents. As suggested
in \onlinecite{MMP00}, the model appears to be
in the same universality class as the
zero-temperature random field Ising model of hysteresis in magnets.
\end{abstract}
\pacs{64.60.Ht,71.45.Lr}

\begin{multicols}{2}      
\section{Introduction} 
Since the eighties the depinning of various {\it elastic} interfaces, 
ranging from domain walls in magnets to geological faults, has received great
attention. Many elastic interface models with dissipative dynamics
display a continuous transition from pinned to sliding at a critical
driving force $F_c$ \cite{fisher85,NF92}. Near this transition the
behavior is universal on long length scales, {\it i.e.} it does not
depend on microscopic details, but
only on general properties of the system,
such as symmetries, dimensions, range of interactions, and the dynamics.
The sliding state is unique and no hysteresis can occur \cite{middleton92}.
This class of
systems has been studied extensively by renormalization group (RG)
methods \cite{NF92} and numerical simulations \cite{AAMthesis}.
Meanwhile, a large body
of experimental \cite{shobo93,henderson98} and numerical work
\cite{shi91,faleski96} has shown that  many extended condensed matter systems
with strong disorder exhibit a spatially inhomogeneous {\it plastic} response,
when set into motion by an external drive \cite{fisher98}.
In this plastic flow regime,
topological defects proliferate and the system is broken up in
fluid-like regions flowing around pinned solid regions. The
elastic restoring forces are replaced by
viscous flow on various scales \cite{coppersmith91}, allowing
for hysteretic response.
Examples include vortex arrays in disordered  superconductors,
charge density waves (CDWs) in anisotropic metals, colloids, and many others.
Only for weak disorder these extended systems can be described as elastic
objects pulled through a quenched random medium by a uniform force $F$.
Although several {\it mean-field} (MF) models of
dissipative dynamics with locally underdamped relaxation have been
proposed in the literature in various physical contexts
\cite{strogatz88,SF01,fisher98} and shown to
exhibit hysteresis, very little is known analytically about
the behavior of such models in finite dimensions.
This article provides first analytical (RG) results for a viscoelastic model
that was introduced and previously studied in MF theory
by Marchetti and collaborators \cite{MMP00},
and makes an important contribution towards the identification
of universality classes for nonequilibrium disordered systems.
In \onlinecite{MMP00} it was shown that in MF theory the model
has a critical point, separating continuous from hysteretic dynamics.
In this paper we show that in finite dimensions
the behavior maps onto critical hysteresis in disordered magnets
as modeled by the zero temperature non-equilibrium random field Ising
model (RFIM) \cite{DS9396}. By showing that the two models
are in fact in the same universality class     
we obtain the critical exponents for hysteresis in plastic flow
in an expansion in $6-\epsilon$ dimensions.

 

\section{The driven viscoelastic model} 
\begin{figure}
\begin{center}
\leavevmode
\epsfxsize=7.5cm
\epsffile{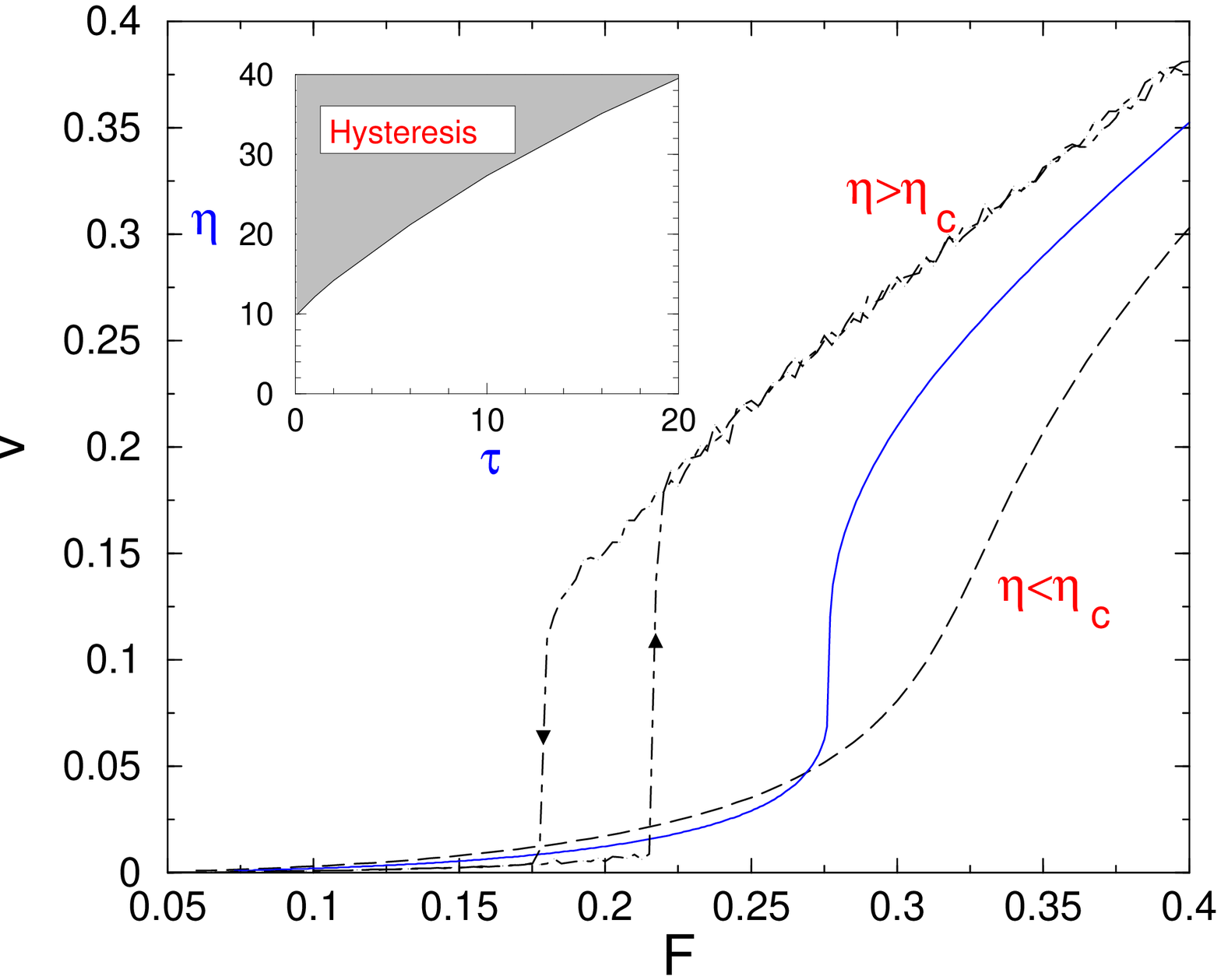}
\end{center}
\caption{\label{figure1} Typical MF $v-F$ curves for 
$\tau=0$ and $\eta=24$
(dashed-dotted), $\eta=\eta_c=9.745$ (solid),
and $\eta=6$ (dashed). 
The inset shows the phase diagram in the $(\eta,\tau)$ plane.
The shaded area is the region where the dynamics is hysteretic.}
\end{figure}

\begin{figure}
\begin{center}
\leavevmode
\epsfxsize=7.5cm
\epsffile{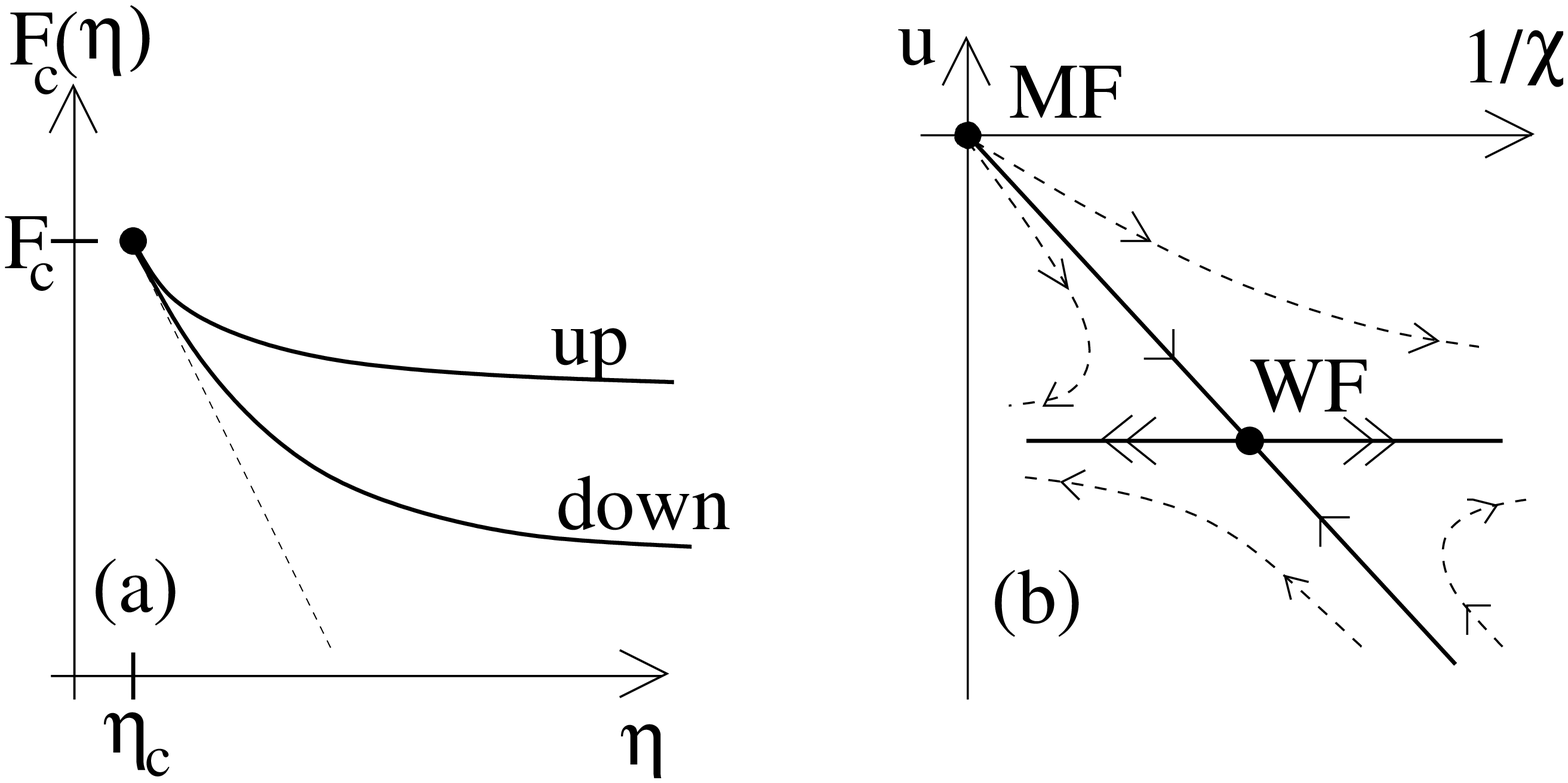}
\end{center}
\caption{\label{figure2} Schematic MF phase diagram and RG flows.
(a) The bold curves are the critical
lines  $F_c(\eta)$ for the onset of the vertical jump in the MF response
({\it i.e.} where $d\overline{v}/dF\rightarrow\infty$)
when the force $F$ is (i) increased adiabatically from a static random 
configuration at $F=0$ 
(up);
and (ii) decreased adiabatically from $+\infty$ (down).
The critical point studied here is at the end of these lines,
$(\eta_c,F_c(\eta_c))$. (b) RG flows below 6 dimensions in the
$(\chi^{-1},u)$ plane. The MF fixed point  becomes unstable
below $d=6$
and the critical behavior is obtained by linearizing around the Wilson
Fisher (WF) fixed point.}
\end{figure}

In Ref.~\onlinecite{MMP00} it is proposed
that a  description
of shear deformations in the plastic regime may be obtained by
focusing on the dynamics of coarse-grained degrees of freedom
(the solid-like regions) that are allowed to slip past each other.
The model of a driven viscoelastic medium is then obtained by replacing
the elastic couplings of displacements
in the coarse-grained model of an overdamped driven elastic medium
with Maxwell-type couplings of velocities \cite{MMP00}.
Considering, for simplicity,  the overdamped dynamics of a scalar field,
the equation of motion for the
local displacement at discrete
lattice sites $i$,
$u({\bf r},t)\rightarrow u_i(t)$,
is
\begin{equation}
\label{viscoel}
\gamma_0\dot{u}_i=\sum_{\langle ij\rangle}\int_0^t
ds\,\mu_{ij}e^{(t-s)/\tau}\big[\dot{u}_j(s)-\dot{u}_i(s)\big]
+F+F_i(u_i),
\end{equation}
where the dot denotes a time derivative, 
the summation is restricted to nearest neighbor pairs and $\gamma_0$
is the friction.
If all the nearest-neighbor elastic
couplings  are equal ($\mu_{ij}=\mu\geq 0$),  the first term on the right hand
side of Eq.\ (\ref{viscoel}) is the discrete Laplacian in $d$
dimensions.  The second term is the homogeneous driving force, $F$,
and $F_i(u_i)$ denotes the pinning force arising from a quenched
random potential, $V_i(u_i)$,
%
$F_i(u_i)=-{dV_i / du_i}=h_i Y(u_i-\beta_i)$,
%
with $Y(u)$ a periodic function with period $1$ and $\beta_i$
random phases uniformly distributed in $[0,1]$. The random pinning forces
$h_i$ are distributed with probability $\rho(h)=e^{-h/h_0}/h_0$,
of width $h_0$. The model can be used to describe pure shear deformations,
where the interactions among the degrees of freedom are 
transverse to the direction of mean motion.
As shown in \onlinecite{MMP00}, the integro-differential equation
can be transformed into a second order differential equation,
given by
\begin{eqnarray}
\label{diff_eqn}
\tau\ddot{u}_i+\Big(\gamma- \tau\frac{\partial F_i}{\partial u_i}\Big)
   \dot{u}_i=
  \sum_{\langle ij\rangle}\eta_{ij}[\dot{u}_j-\dot{u}_i]+ F+F_i(u_i),
\end{eqnarray}
with $\eta_{ij}=\mu_{ij}\tau$. The MF critical line $\eta_c(\tau)$ separating
continuous from hysteretic dynamics was obtained in \onlinecite{MMP00}
and is shown in the inset of Fig. \ref{figure1}.
\noindent In this paper we consider the limit
$\tau=0$, with $\eta=\mu\tau\not=0$, referred to as the viscous limit,
and study the critical point
along the $\tau=0$ line. 
The MF analysis suggests that that any finite value of
$\tau$ may be irrelevant and the large scale behavior be generically
described by $\tau=0$. In this limit the model reduces to
\begin{eqnarray}
\label{model}
\dot{u}_i=  
  \sum_{\langle ij\rangle}\eta_{ij}[\dot{u}_j-\dot{u}_i]+ F+F_i(u_i).
\end{eqnarray}
We have introduced dimensionless variables by incorporating $\gamma_0$ in our unit
of time and scaling all forces and velocities by the width $h_0$ of the disorder distribution.
The relevant dimensionless tuning parameters are then the strength $\eta$
of the viscous coupling and the driving force, $F$.
The MF solution of Eq. (\ref{model}) for piece-wise parabolic 
pinning potential was obtained in \cite{MMP00}
and is given by
\begin{equation}
\overline{v}=\Big\langle\frac{1}{T(h_i)}\Big\rangle
\end{equation}
where $\langle ...\rangle$ denotes an average over disorder and
$T(h_i)$ is the period of the $i$-th degree of freedom ({\it i.e.} the time
over which
the displacement $u_i$ is incremented by one),
\begin{eqnarray}
\label{period}
& & T(h)=\frac{1+\eta}{h}\Big[\ln\Big(\frac{2g+h}{2g-h}\Big)\Big]\;,\hspace{0.2in} {\rm for}\hspace{0.2in} h\leq 2g\;,\nonumber\\
& & T(h)=\infty\;, \hspace{0.2in} {\rm for}\hspace{0.2in} h> 2g\;,
\end{eqnarray}
with $g=F+\eta\overline{v}$.
The various types of dynamical response are shown in Fig. 1.
For strong coupling (relative to disorder), $\eta>\eta_c$,
a marginally pinned degree
of freedom pushes over its neighbors, driving a 
vertical hysteretic jump in the $\overline{v}-F$
curve between
strongly pinned (slowly moving) and weakly pinned (fast
moving) states.
For weak coupling ($\eta<\eta_c$)
most degrees of freedom advance independently
and the $\overline{v}-F$ curve has no macroscopic jumps. 
In MF the hysteresis disappears at (and below) a special 
value $\eta=\eta_c$, where 
the $\overline{v}-F$ curve has a vertical slope.
Preliminary results indicate the same for simulations 
in three dimensions (up to fluctuations).
Figure \ref{figure2} shows a schematic phase diagram for the dynamical model
defined by Eq. (\ref{model}) in the $(F,\eta)$ plane.
The MF critical point $(F_c,\eta_c)$ is obtained
from
$\chi=d\overline{v}/dF\rightarrow\infty$
and  $d^2\overline{v}/dF^2\rightarrow\infty$,
where  $\chi$ is the static response function.
As shown in Appendix A, near this critical point
the velocity $\overline{v}-v_c$, with
$v_c=\overline{v}(\eta_c,F_c)$,         
obeys a universal scaling law,
\begin{equation}
\label{scaling_fct}
\overline{v}-v_c\sim |r|^\beta {\cal G}_{\pm}\Big(\frac
{f'}{|r|^{\beta\delta}}\Big),
\end{equation}
where
$r=(\eta-\eta_c)/\eta_c$
and $f'=f\pm v_c\eta_c |r|/(1+\eta_c)$,
with $f=F-F_c(\eta_c)$, is a (non-universal)
rotation between the control variables $(r,f)$ and
the scaling variables $(r,f')$.
The $\pm$ refers to the sign of $r$.
In MF theory the critical exponents have the values
$\beta_{MF}=1/2$ and $\delta_{MF}=3$.

\section{Beyond Mean-Field Theory}
\subsection{Formalism}
 
Here we set up the formalism to study fluctuations about MF theory. 
We use the method of Martin, Siggia and Rose (MSR) to transform the
stochastic equation of motion (\ref{model})        
into a field theory. This is done by introducing a generating functional
$Z$, which contains the sum over the probabilities of all possible
{\it paths} ({\it i.e.} microscopic configurations) 
which the system follows 
for different configurations  of disorder
as the force $F$ is slowly increased. The generating functional
is written as the exponential of an action, which is then renormalized
perturbatively by standard methods.
The MSR formalism was first adapted to driven
systems by Narayan
and Fisher to describe the depinning of an elastic model of
CDWs \cite{NF92}. It was later employed
by Dahmen and Sethna to study the hysteretic response of
disordered magnets \cite{DS9396}. Here we only give a very
sketchy summary of the method and refer to the literature for details.
The crucial point is to recognize that
the coarse grained local
velocities ($v_i$ defined below) in the present model play the role of
local magnetizations in the RFIM. The formal manipulations then follow closely those of Ref.~\onlinecite{DS9396}, although important differences
occur in the detailed calculation of the response and correlation functions.

The functional $Z$ is defined
as a product of $\delta$-functions (one for each site), each of
which imposes that the dynamics of each $u_i$ is described at all times
by the equation of motion (\ref{model}).
The $\delta$-functions are
then rewritten using a standard identity, so that
\begin{equation}
\label{Z}
1\equiv Z=\int[du][d\hat{u}]J[u]e^S\;,
\end{equation}
with $\hat{u}_i$ MSR auxiliary fields.
Here $J[u]$ is a functional Jacobian chosen so that $Z$ integrates to unity
and $S$ is an action given by
\begin{equation}
\label{action}
S=i\int_{t}\hat{u}_i\Big[\dot{u}_i
  -\sum_{\langle ij\rangle}\eta_{ij}[\dot{u}_j-\dot{u}_i]- F-F_i(u_i)\Big].
\end{equation}   
To study perturbatively fluctuations about MF theory
we change variables from the $u_i$ and $\hat{u}_i$
to the local velocities $v_i=(1/\eta)\sum_j\eta_{ij}\dot{u}_j$ by
introducing another auxiliary field, $\hat{v}_i$.
%
%
%
%
%
The MF solution $(v_0,\hat{v}_0)$
is the saddle point of the resulting generating functional.
By expanding about this solution and shifting the definition of $v$
to $v-v_0$, (and absorbing a factor $i$ in the definition),
we obtain the generating functional, \cite{NF92,DS9396}
\begin{equation}
\label{Zbareff}
\overline{Z} = \int [dv][d\hat{v}]e^{S_{\rm eff}}
\end{equation}     
with an effective action
\end{multicols}
\begin{eqnarray}
\label{Seff}
S_{\rm eff}=-\sum_{ij}\int_t dt \eta^{-1}_{ij}\hat{v}_i(t)v_j(t)
  + \sum_j\sum_{m,n=0}^{\infty} \frac{1}{m!n!}\int_{t_1}...\int_{t_{m+n}}
   u_{mn}(t_1,...,t_{m+n})
 \hat{v}_j(t_1)...\hat{v}_j(t_m)
   v_j(t_{m+1})...v_j(t_{m+n}).
\end{eqnarray}
%
\begin{multicols}{2}
\noindent
The $u_{mn}$ are
the linear and nonlinear {\it local} response and
connected ($\rangle_c$) correlation functions, of the form
\begin{equation}
\label{umn}
u_{mn}=\frac{\partial}{\partial\epsilon(t_{m+1})}...
   \frac{\partial}{\partial\epsilon(t_{m+n})}
   \langle\dot{u}(t_1)...\dot{u}(t_m)\rangle_{l,c}.
\end{equation}
The suffix $l$ in Eq. (\ref{umn}) indicates that the  response and
correlation functions are obtained by solving the MF equation
with fixed {\it local} fields $v_{0}$ and an infinitesimal
perturbation,  $\eta\epsilon(t,t_1)$, at $t_1<t$,
\begin{equation}
\label{perturbedeqn}
(1+\eta)\partial_tu_j(t)=F+\eta v_{0j}+F_j(u_j)+\eta\epsilon(t,t_1).
\end{equation}
The effective action (\ref{Seff}) 
has the same structure as that of the spin model in Ref.~\cite{DS9396},
with local velocities $v_i$ replacing local magnetizations.
Note that while in the RFIM at fixed field all spins are fixed in
time (in the steady state), in the CDW model each phase is subject to a
quasiperiodic noise due to its coupling to the neighboring sites 
that move in a periodic potential even at fixed force. In the calculation
done here this additional noise is averaged out and does not seem to affect
the scaling results on long length scales. A detailed study whether or not 
it matters on long length scales in ways not captured by this
approach is left to future work. 
  
\subsection{Results}
The response function of order $n$ is calculated by perturbing the system   
with $n$ $\delta$-function pulses of strength
$\epsilon_1,\epsilon_2,...,\epsilon_n$
at $t_1<t_2<...<t_n$,  and evaluating the response at a time $t>t_n$.
In contrast to the case of driven CDWs \cite{NF92}, terms of order
$\sim e^{-\lambda T(h)}$, with $\lambda=h/(1+\eta)$,
cannot be neglected in the calculation, as, even when the mean
velocity is very small, a fraction of degrees of freedom may be sliding
freely, with corresponding large periods. Furthermore,
the response must be evaluated over the entire history to obtain
the correct low frequency behavior.
An example of such a calculation for the simplest case of the linear
response function is shown in Appendix B.
Although the details of
the calculation for the present model are different from
the spin model, we find that
the $u_{mn}$ of the two models have indeed the same long wavelength 
and low frequency
behavior, {\it i.e.} the effective actions of the two models
are identical. Consequently, the RG analysis of the
viscoelastic model is the same as for the RFIM  \cite{DS9396}
and is not reproduced here. 
(We obtain the same $\epsilon$-expansion to all orders in $\epsilon$.
Note that while the 
$\epsilon$-expansion for the {\it equilibrium} RFIM is controversial 
\cite{Feldman},
and the $\epsilon$-expansion for the nonequilibrium RFIM has been mapped
onto that of the equilibrium RFIM \cite{DS9396} to all orders in 
$\epsilon$, our main result here
does not depend on the outcome of this controversy: the mapping of 
the plastic CDW depinning model onto the hysteretic RFIM is based on
mapping the actions of the two models onto each other, by identifying
the local velocities in the CDW model with the local magnetizations 
in the spin model. This mapping holds
independently of the subsequent $\epsilon$- expansions
for the respective critical exponents.)
The critical exponents of the two systems are the same
and are given in Table 1 \cite{noise}.
Fourier transforming the fields $v$ and $\hat{v}$ in both space and time,
the quadratic part of the effective action is given by
%
\begin{eqnarray}
S_{\rm eff}^G=&-&\int_{{\bf q},\omega}\hat{v}(-{\bf q},-\omega)
  \big[\eta\eta^{-1}(q)-u_{1,1}(\omega)\big]v({\bf q},\omega)\nonumber\\
  & & +\frac{1}{2}\int_{{\bf q},\omega}\hat{v}(-{\bf q},-\omega)
   u_{2,0}(\omega)\hat{v}({\bf q},\omega),
\end{eqnarray}
%
where  $u_{1,0}$ is trivially zero because we expand
around the stationary point.
In the long wavelength and low frequency limit, we approximate
$\eta^{-1}(q)\approx\frac{1}{\eta}+Kq^2$ and
$u_{1,1}(\omega)\approx u_{1,1}^{\rm stat}+i\omega a$,
where $u_{1,1}^{\rm stat}=\eta\nu'(g_0)/(1+\eta)$,
with $g_0=\eta v_0+F$, and $u_{1,1}^{\rm stat}$ and $a$ are given in Appendix B.
The correlation function is 
$u_{2,0}(\omega)\approx 2\pi\delta(\omega)\Big[\langle\big(\frac{1}{T}\big)^2\rangle_h-\langle\frac{1}{T}\rangle_h^2\Big]$.
The bare propagator is given by
$G_{\hat{v}v}(q,\omega)\approx [-i\omega a+\eta Kq^2-\chi^{-1}]$,
where $\chi=[u_{1,1}^{\rm stat}-1]^{-1}$ is precisely
the static response to a
monotonically increasing external force calculated in MF theory.
At the MF critical point $\chi^{-1}=0$ and
the bare propagator becomes diffusive,
$G_{\hat{v}v}(q,\omega=0)\sim 1/q^2$, while the correlation
function is static, with
$G_{vv}(q,\omega)\sim\delta(\omega)/q^4$.
This behavior is the same as for the RFIM. In contrast, the bare propagator
for driven {\it elastic} CDWs is diffusive also away from the critical point.
 
To set up an RG calculation, we need
to consider terms beyond Gaussian in the action.
To obtain an approximate effective action,
we Fourier transform the fields in time, retain only the low frequency
limit of the response functions,  and then transform back to time.
In the static limit, the $u_{1,2}$ term has bare value
$u_{1,2}(\omega_1,\omega_2)\approx w$, with
$w \sim \eta^2 \nu''(g_0)/(1+\eta)$, and
$u_{1,3}(\omega_1,\omega_2,\omega_3)\approx u=\eta^3 \nu'''(g_0)/(1+\eta)$.
Note that $w=0$ at the MF
critical point.
We then obtain
%
%
\begin{eqnarray}
\label{Seff_tot}
S_{\rm eff}\approx S_{\rm eff}^G
      +\sum_i\int_t\Big\{\frac{w}{2}~\hat{v}_i\big[v_i(t)\big]^2
      +\frac{u}{6} ~\hat{v}_i(t)[v_i(t)]^3\Big\},
\end{eqnarray}
%
\noindent
which is identical to the effective action for the soft spin RFIM \cite{DS9396}, 
with the local velocities here playing the role of the local magnetizations
in the RFIM.
The generic behavior at long time and length scales is therefore identical to
that of the RFIM. To perform the coarse-graining transformation, as usual we
integrate out modes in the wave vector shell $\Lambda/b<q<\Lambda$ ($b>1$),
rescale coordinates as $x=bx'$ and $t=b^zt'$ and choose the rescaling of the
fields so that the quadratic part of the action is unchanged at the critical
point ($\chi^{-1}=0$). This requires $z=2$ and yields
$u'_{mn}=b^{[-(m+n)+2]d/2+2n}u_{mn}$.
 
If $d>8$, all $u_{mn}$ other than those yielding quadratic terms in the
action renormalize to zero and become irrelevant at large scales,
indicating that the MF line for the onset
of the vertical jump in the $\overline{v}-F$ curve
(where $\chi^{-1}=0$ and $w\not=0$) remains critical
\cite{noise}. The MF scaling exponents are exact in this case.

\vspace*{0.1in}
\begin{center}
\begin{tabular}{|c|c|c|}\hline
Exponents & Mean-field & $\epsilon$ expansion\\\hline\hline
$\beta$ & $1/2$ & $1/2-\epsilon/6+{\cal O}(\epsilon^2)$ \\\hline
$1/\nu$ & $2$ & $2-\epsilon/3+{\cal O}(\epsilon^2)$ \\\hline
$\beta\delta$ & $3/2$ & $3/2+0.0833454\epsilon^2$ \\\hline
\end{tabular}
\end{center}
 
\vspace{0.1in}
\noindent{\small TABLE I: Universal exponents for the critical
point discussed in the text. The exponents $\beta$ and $\delta$ describe how
$\overline{v}$ scales with $F$ and $\eta$, respectively.
$\nu$ is the correlation length exponent.}
%
\vspace{0.1in}    
 
For $d<8$ we need to distinguish two cases, depending on whether
$w'\sim b^{4-d/2}w$ is initially finite or zero.
A bare action with $\chi^{-1}=0$ and $w\not=0$ at the onset line $F_c(\eta)$
corresponds to the hysteretic region $\eta>\eta_c$. In this case the RG
analysis carried
out by Dahmen and Sethna for the RFIM shows that 
$w$ flows to $\infty$, suggesting a sharp jump onset in the hysteresis loop
for $\eta>\eta_c$.
A bare action with $\chi^{-1}=0$ and $w=0$
corresponds to 
the critical point of interest here. If the bare $w$ is zero,
we must consider higher order terms. In particular we find that
$u'\sim b^{6-d}u$, while all the higher order response
functions are irrelevant for $d=6-\epsilon$. So $d_c=6$ is the upper critical dimension
and for $d>6$ the critical exponents near $(F_c,\eta_c)$ are given by MF theory.
Below 6 dimensions, $u$ is relevant. Corrections to the critical
exponents were computed in  $6-\epsilon$ dimensions in \cite{DS9396},
the results are summarized in Table 1.
%
%

Many open questions remain. First, the RG calculation 
near $(F_c,\eta_c)$ described here
applies for $\eta<\eta_c$ \cite{DS9396}. More work is needed 
to extend it above $\eta_c$, as well as for a proper
understanding of the noise in this region.  Secondly, we would like to study
the finite range viscoelastic model for general value of $\tau$.
\vspace{0.2in}
 
We thank Alan Middleton, Jennifer
Schwarz for illuminating discussions and help with one 
of the figures. We also thank Jim Sethna for helpful comments.
MCM was supported by NSF through grants DMR97-30678 at Syracuse and
PHY99-07949 at ITP, Santa Barbara.
KAD acknowledges support from the Materials Computation Center
(grant NSF-DMR 99-76550), NSF grant DMR00-72783, and an A. P. Sloan
fellowship.    

\section{Appendix A: Mean-Field Scaling}
Here we derive the scaling behavior near the MF critical point.
It is convenient to rewrite the mean velocity as
\begin{equation}
\label{def_nug}
\overline{v}= \frac{\nu(g)}{1+\eta}\;,
\end{equation}
by introducing the function $\nu(g)$,
\begin{equation}
\label{nug}
\nu(g)=\int_0^{2g}dh~\rho(h)~\frac{h}{\ln\Big(\frac{2g+h}{2g-h}\Big)}\;,
\end{equation}
with $\rho(h)$ the normalized distribution of pinning strengths.
The MF critical point, $\eta=\eta_c$, $F=F_c$ and $\overline{v}=v_c$,
is the point where 
the mean velocity curve has infinite slope. 
The conditions for
the critical point easily are expressed in terms $\nu(g)$ 
by differentiating both sides
of Eq.~(\ref{def_nug}) with respect to $g$, to obtain
\begin{eqnarray}
\label{nugp}
& & \Big(\frac{\partial \overline{v}}{\partial F}\Big)_{\eta}\rightarrow\infty
\hspace*{0.2in}\Rightarrow\hspace*{0.2in}
\nu'(g)=\frac{1+\eta}{\eta}\;,\\
\label{nugpp}
& &\Big(\frac{\partial^2 \overline{v}}{\partial F^2}\Big)_{\eta}\rightarrow\infty
\hspace*{0.2in}\Rightarrow\hspace*{0.2in}\nu''(g)=0\;,
\end{eqnarray}
where the prime denotes a derivative with respect to $g$.

To determine the critical line $F_c(\eta)$ where the slope of the
$(\overline{v},F)$ curve diverges, we solve Eqs.~(\ref{def_nug}),
(\ref{nugp}) and (\ref{nugpp})
for $\eta$ near $\eta_c$.
By expanding both sides of Eq.~ (\ref{def_nug})
in  $\delta g_c(\eta)=g_c(\eta)-g_c$ and $\delta\eta=\eta-\eta_c$, 
and using that $\nu''(g_c)=0$, we obtain 
$\delta g_c(\eta)\sim \pm \Big[\frac{-2\delta\eta}{\eta_c^2\nu_c'''}\Big]^{1/2}$,
with $\nu_c'''=\nu'''(g_c)<0$. 
We then use $\delta F_c(\eta)=\delta g_c(\eta)
-v_c\delta\eta-\eta_c\delta\overline{v}_c
-\delta\eta\delta\overline{v}_c$, expand both sides of Eq. (\ref{def_nug}) 
and eliminate $\delta g_c(\eta)$ between the two equations,
to obtain
\begin{equation}
\delta F_c(\eta)=-\frac{v_c}{1+\eta_c}\delta\eta
   \pm\frac{2}{3\eta_c^2(1+\eta_c)}\Big[\frac{-2}{\nu'''_c}\Big]^{1/2}
   (\delta\eta)^{3/2}\;.
\end{equation}
Note that the critical line $F_c(\eta)$ is rotated as compared to
the critical line in the magnetic system. 

To obtain the scaling function, we now write
$\overline{v}(F,\eta)=\overline{v}(F_c,\eta_c)+\delta\overline{v}$
in Eq. (\ref{def_nug}), expand around the critical point and eliminate $\delta g$
in favor of $\delta\overline{v}$, $\delta\eta$ and $\delta F$.
Retaining only terms up to cubic in the deviations from the critical point,
we obtain a cubic equation
for $\delta\overline{v}$, given by
\begin{equation}
\label{cubic1}
\frac{\eta_c^4\nu_c'''}{3!(1+\eta_c)}(\delta\overline{v})^3
  +\frac{1}{1+\eta_c}\delta\eta\delta\overline{v}
  +\frac{v_c}{1+\eta_c}\delta\eta
  +\delta F =0\;.
\end{equation}
We define a rotation from $(F,\eta)$ to new scaling variables
$(F',\eta)$, with $F'=F+\frac{v_c}{1+\eta_c}\eta$.
In terms of the new variables Eq.~(\ref{cubic1}) becomes
\begin{equation}
\label{cubic2}
(\delta\overline{v})^3
  +\frac{6}{\eta_c^4\nu_c'''}\delta\eta\delta\overline{v}
  +\frac{6(1+\eta_c)}{\eta_c^4\nu_c'''}\delta F' =0\;,
\end{equation}
whose solution is given by Eq. (\ref{scaling_fct}), with ${\cal G}_\pm(x)$
the smallest real root of the cubic equation
\begin{equation}
\label{cubic3}
{\cal G}_\pm^3\pm\frac{6}{\eta_c^3\nu_c'''}{\cal G}_\pm
  +\frac{6(1+\eta_c)}{\eta_c^4\nu_c'''}x=0\;.
\end{equation}
The scaling function satisfies
${\cal G}_\pm(0)={\rm constant}$ and ${\cal G}_\pm (x>>1)\sim x^{1/3}$.

\section{Appendix B: Evaluation of the linear response}
The linear response function $u_{1,1}$ is obtained by perturbing the 
system with a $\delta$-function pulse of strength $\epsilon>0$ at time $t_1$
and evaluating the response $\langle\dot{u}(t)\rangle_{l,c}$ 
at time $t>t_1$,
\begin{equation}
u_{1,1}(t,t_1)=
\frac{\partial\langle\dot{u}(t)\rangle_{l,c}}{\partial\epsilon(t_1)}\;,
\end{equation}
where $u(t)$ is obtained by solving Eq.~(\ref{perturbedeqn}) for 
$\epsilon(t,t_1)=\epsilon\delta(t-t_1)$ (to simplify the notation, we drop 
the subscript $0$ on the local field $v_0$ in Eq.~(\ref{perturbedeqn})).
The other local response and correlation functions can be evaluated 
by similar methods.

\begin{figure}
\begin{center}
\leavevmode
\epsfxsize=6.5cm
\epsffile{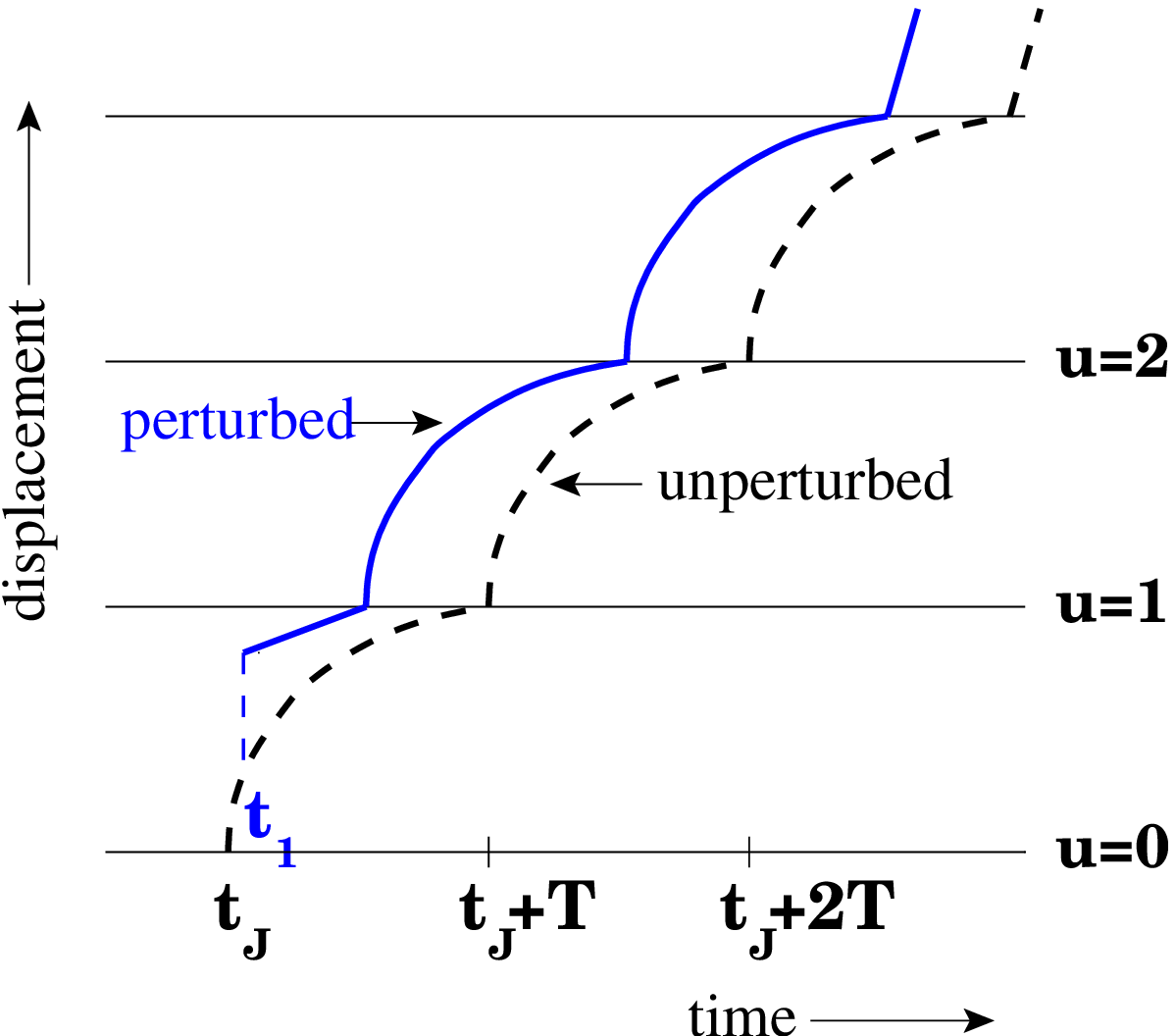}
\end{center}
\caption{\label{phase} Time evolution of the displacement $u(t)$. Both
the unperturbed
(dashed) and perturbed (solid line) solutions are shown.
The perturbed solution is obtained by applying a $\delta$-function pulse 
at time $t_1$,
with $t_J\leq t_1\leq t_1+T$.}
\end{figure}

As shown in Fig.~\ref{phase}, the perturbation has two effects: it yields 
a discontinuous jump of the displacement at time $t_1$ and
it causes a shift in the ``jump time'' $t_J$ where the displacement jumps
between neighboring wells of the periodic pinning potential.
Without loss of generality we assume $t_J\leq t_1\leq t_J+T$,
where $T$ is the period defined in Eq.~(\ref{period}) and $t_J$ is the jump time 
of the unperturbed solution. 

For $t\geq t_1$, the perturbed solution can  be written as
\begin{equation}
u(t,t_1;t_J)=\Theta(t-t_1)u_{\rm unp}(t+\delta t_J;t_J)\;,
\end{equation}
where  $u_{\rm unp}(t;t_J)$ is the unperturbed solution,
given by
\end{multicols}
\begin{equation}
u_{\rm unp}(t;t_J)=
  \sum_{n=0}^\infty\Theta(t_J+(n+1)T-t)\Theta(t-t_J-nT)
   \Big[n+\frac{1-e^{-\lambda(t-t_J-nT)}}{1-e^{-\lambda T}}\Big]\;,
\end{equation}
\begin{multicols}{2}
\noindent with $\lambda=h/(1+\eta)$.
The shift in jump time $\delta t_J$ is determined by requiring
$u(t_J-\delta t_J,t_1;t_J)=0$,
with the result,
\begin{eqnarray}
\delta t_J =-\frac{1}{\lambda}\ln\big(1-\epsilon\alpha
  e^{-\lambda(t_J+T-t_1)}\big)\;,
\end{eqnarray}
if $t_J\geq t_1-T+t_\epsilon$, with $t_\epsilon=\frac{1}{\lambda}\ln(1+\epsilon\alpha)$, 
and $\delta t_J=0$ otherwise,
with $\alpha=\eta(e^{\lambda T}-1)/(1+\eta)$.

The local response is obtained by averaging over a uniform 
distribution of jump times,
as well as over the distribution $\rho(h)$ of pinning strengths.
The average over jump times of the perturbed velocity is given by
\end{multicols}
\begin{equation}
\label{integral}
\langle \dot{u}\rangle_{t_J}=\int_{t_1-T}^{t_1-T+t_\epsilon}\frac{dt_1}{T}~
 \frac{\partial}{\partial t}\Big[\Theta(t-t_1)u_{\rm unp}(t;t_J)\Big]
  +\int_{t_1-T+t_\epsilon}^{t_1}\frac{dt_1}{T}~
 \frac{\partial}{\partial t}\Big[\Theta(t-t_1)u_{\rm unp}(t+\delta t_J;t_J)\Big]\;.
\end{equation}
%
Carrying out the integrals \cite{foot_integral}, differentiating the response with
respect to $\epsilon$ and evaluating the result at $\epsilon=0$,
we obtain the mean {\it local} linear response function,
$\chi_l(t-t_1)=u_{1,1}(t,t_1)$,
with
%
\begin{equation}
\chi_l(t)=\delta(t)\frac{\eta}{1+\eta}
  +\Theta(t)\Big\langle\frac{\eta\lambda}{(1+\eta)T}\sum_{n=0}^\infty
    \Theta(t-nT)\Theta((n+1)T-t)e^{-\lambda(t-nT)}
     \Big[\Big(nT-t+\frac{1}{\lambda T}\Big)(e^{\lambda T}-1)-1\Big]\Big\rangle\;,
\end{equation}
\begin{multicols}{2}
\noindent where the brackets $\langle ...\rangle$ denote the average over $\rho(h)$.
For $t\geq 0$, the response consists of the sum of a $\delta$-function contribution and 
a periodic function. The frequency dependent local response is given by
\begin{eqnarray}
\tilde{\chi}_l(\omega)&=&\int_0^\infty dt~e^{i\omega t}\chi_l(t)\;,\nonumber\\
&=&\frac{\eta}{1+\eta}
   +\Big\langle\frac{1}{T}\sum_s\frac{g_s}{\zeta-i(\omega-\omega_s)}\Big\rangle\;,
\end{eqnarray}
where the limit $\zeta\rightarrow 0$ is intended, $\omega_s=2\pi s/T$, with
$s$ an integer, and
\begin{equation}
g_s=-\frac{\eta\lambda e^{\lambda T}}{1+\eta}
     \frac{(1-e^{-\lambda T})^2}{\lambda T}
     \frac{i\omega_s}{(i\omega_s-\lambda)^2}\;.
\end{equation}
For small frequency, we can carry out the summation over $s$ and find
\begin{equation}
\tilde{\chi}_l(\omega)\approx u_{1,1}^{\rm stat}+i\omega a\;,
\end{equation}
with
\begin{eqnarray}
u_{1,1}^{\rm stat}=\frac{\eta\nu'(g)}{1+\eta}\;,
\end{eqnarray}
and
\begin{eqnarray}
a=& &-\frac{\eta}{1+\eta}\Big\langle
  \frac{1}{\lambda}+\frac{1}{2\lambda^2T}(e^{\lambda T}-e^{-\lambda T})\nonumber\\
 & &   -\frac{2}{\lambda^3T^2}(e^{\lambda T}-1)(1-e^{-\lambda T})\Big\rangle\;.
\end{eqnarray}

\end{multicols}
 

\begin{references}
\bibitem{foot_integral}
To evaluate the second integral in Eq.~(\ref{integral})
it is convenient to introduce the
nonlinear change of variable $y=t_J-\delta t_J(t_J)$.
\end{references}
\end{document}